# Scrape-off layer ion temperature measurements at the divertor target during type III and type I ELMs in MAST measured by RFEA

**S. Elmore**[a], S. Y. Allan[a], G. Fishpool[a], A. Kirk[a], A. J. Thornton[a], N. R. Walkden[a], J. R. Harrison[a] and the MAST Team[a]

[a]CCFE, Culham Science Centre, Abingdon, Oxon, OX14 3DB, UK

## Abstract

In future nuclear fusion reactors high heat load events, such as edge-localised modes (ELMs), can potentially damage divertor materials and release impurities into the main plasma, limiting plasma performance. The most difficult to handle are type I ELMs since they carry the largest fraction of energy from the plasma and therefore deposit the largest heat flux at the target and on first wall materials. Knowing the temperature of the ions released from ELM events is important since it determines the potential sputtering they would cause from plasma facing materials. To make measurements of $T_i$ by retarding field energy analyser (RFEA) during type I ELMs a new operational technique has been used to allow faster measurements to be made; this is called the fast swept technique (FST). The FST method allows measurements to be made within the time of the ELM event which has previously not been possible with $T_i$ measurements. This new technique has been validated by comparing it with a slower average measurement previously used to make ion temperature measurements of ELMs. Presented here are the first $T_i$ measurements during Type I ELMs made at a tokamak divertor. Temperatures as high as 20 eV are measured more than 15 cm from the peak heat flux of an ELM, in a region where no inter-ELM current is measured by the RFEA; showing that ELM events cause hot ions to reach the divertor target far into the scrape off layer. Fast camera imaging has been used to investigate the type of ELM filaments that have been measured by the divertor RFEA. It is postulated that most of the ion temperatures measured in type I ELMs are from secondary ELM filaments which have not been previously identified in MAST plasmas.

# 1 Introduction

A significant challenge in the design of future nuclear fusion reactors is handling the power exhaust from off-normal events such as edge-localised modes (ELMs) [1]. Previous measurements made in JET [2], ASDEX Upgrade (AUG) [3] [4]and MAST [5] with retarding field energy analysers (RFEA) have shown that ELMs can carry significant amounts of energetic ions into the far SOL at the mid-plane. For tokamaks with close fitting vessel walls, such as ITER, this can be a significant problem since high thermal loading to first wall and divertor materials may cause damage. Furthermore impurities can be sputtered from first wall materials by high energy ion impact and then enter the main plasma and degrade performance.

Measuring the ions released from ELMs which arrive at the divertor target can give information about the motion of ELM filaments and the evolution of the ion energies. Although this is challenging due to the fast timescales involved, ions are ideal to measure since they lose their energy slower than electrons and therefore are a tool to measure where the ELM energy is deposited.

Previously, measurements of ion temperatures have been made by RFEA which give an average or maximum temperature for the ELM ions. These measurement techniques, which have been used on MAST at the mid-plane [5] [6] and at the divertor [7], operate on timescales slower than the ELM event and therefore do not give any information on the evolution of the temperature during the ELM rise and fall as seen in both deuterium alpha ($D_\alpha$) light at the mid-plane and current signals measured by electrical probes such as Langmuir probes (LP). A new measurement technique and associated analysis method has been employed on MAST using the divertor RFEA which allows measurements within the time during which ions from the ELM arrive at the target.

The measurements focus on Type I ELMs which have been seen to deposit the most energy at the target of a tokamak [8]. The RFEA measures far from the last closed flux surface (LCFS) at the target, $\Delta R_{LCFS}^{tgt} > 15$ cm, at the target which allows investigation into ion temperatures corresponding to ELM events in a region far from the strike point; which typically sees the highest temperatures and heat fluxes. This will allow an understanding of how far hot ELM ions can reach. It has already been seen on MAST that type I ELMs show high ion temperatures (~30 eV) existing at the mid-plane up to 16 cm from the LCFS [6] and ELM ions with energy exceeding 500 eV exist 19 cm away from the LCFS [5].

The work presented here builds on the first measurements made of type III ELMs at the divertor in MAST [9] by measuring type III and type I ELMs with the new fast sweeping technique. Using measurements of type III ELMs in a similar scenario to those in [9] the validity of the technique has been confirmed before it has been applied to several type I ELMs. Type I ELMs are studied individually to find how $T_i$ varies in time. The RFEA signals measured during type I ELMs have been found in radial regions of the divertor, $\Delta R_{LCFS}^{tgt} > 15$ cm, not experiencing significant heat flux steady state or measuring any ion current in inter-ELM periods. The signals seen in type I ELMs exist over longer timescales than is usually seen for type I ELMs on MAST. The reason postulated for this, explored in this paper, is the presence of secondary ELM filaments [10] which will be investigated further in section 3.

The RFEA ELM measurements are compared with infra-red (IR) [11] thermography measurements at the outer divertor target of MAST, observations of ELM filaments using visible imaging [12] [13] and the Ball Pen Probe (BPP) [14] installed on the mid-plane reciprocating probe (RP).

## 2 Experimental Technique

Measurements of the ion temperature have been made at the outer lower divertor of MAST [15] using the divertor RFEA probe [7] installed on the divertor science facility (DSF) [16]. The RFEA samples the parallel ion velocity distribution and is therefore angled to align with the field at the divertor target. Further detailed information about the DSF RFEA probe can be found in [7].

Type III ELMs have been measured to confirm both the previous measurement technique [9] and the new measurement technique are in agreement. Following this type I ELM measurements have been studied. Type III ELMs are measured in double-null (DN) discharges with plasma current $I_p$ = 600 kA at a beam power of $P_{NBI}$ = 1.5 MW; which is similar to those studied in [9]. As with previous measurements, the strike point has been held at a constant position during the ELMy H-mode [9], with a radial position of $\Delta R_{LCFS}^{tgt}$ = 5.5 ± 0.5 cm which is defined by the distance between the RFEA probe and the peak heat flux as measured by IR thermography at the lower divertor [11]. Measurements of type I ELMs have been made in several lower single-null (LSN) discharges with $I_p$ = 600 kA and neutral beam power of $P_{NBI}$ = 3.5 MW. As previously mentioned the measurements of type I ELMs are made at large radius to investigate whether hot ions exist in the region $\Delta R_{LCFS}^{tgt}$ > 15 cm. Figure 1 shows the equilibrium for a DN and LSN plasma and the physical location of the RFEA probe installed at the DSF. The closer proximity of PFCs should be noted in the LSN, specifically the relative position of the plasma to the lower P3 coil.

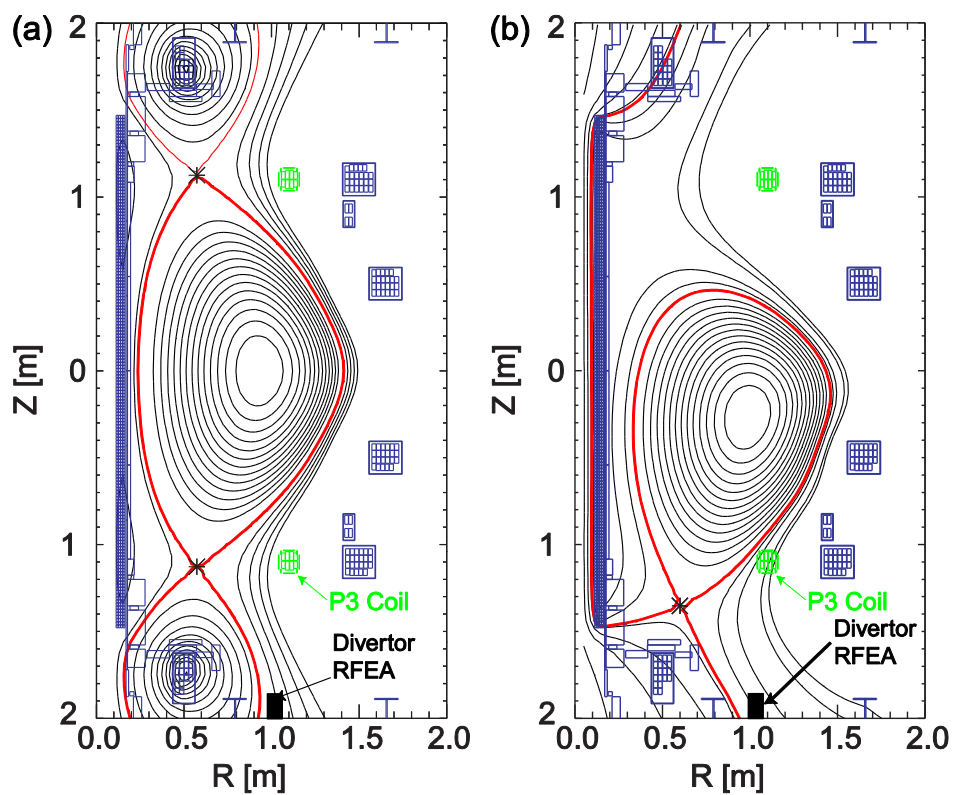

**Figure 1** Poloidal cross section of MAST showing the location of the divertor RFEA and the location of the P3 coil structure in (a) a double null plasma and (b) a lower single null plasma. The P3 coil is much closer to the plasma in (b).

Two operational methods can be used to measure ELM ion temperatures using RFEAs. The technique used previously to measure type III ELMs [9] and L-mode filaments [17] is the slow swept technique (SST). This only allows measurements of an average ELM or filament since it samples a number of similar ELMs (or filaments), each with a different voltage applied to the discriminating grid, see below. The new method, used for the first time on the DSF RFEA at the MAST lower divertor, is the fast swept technique (FST) where the voltage on the discriminating grid is swept at a much higher rate (10 kHz) than normal steady state operation (1 kHz) and can give several RFEA measurements during the time of an ELM arriving at the target; a duration of between 0.6 and 2 ms. Time dependant measurements are made as a function of $t-t_{ELM}$, where $t_{ELM}$ is defined as the time at which the deuterium alpha emission from the mid-plane ($D_\alpha$) rises to 50% of the peak $D_\alpha$ value above the background $D_\alpha$, similar to the criteria in [18].

## 2.1 Fast Swept Technique (FST)

In this use of the FST, triangular sweeps of grid 1 voltage are applied between 0 V and +190 V at 10 kHz. The complete waveform, returning to 0V, has a period of 100 μs, however to maximise time resolution, rising and falling sweeps of discriminating voltage are treated separately here, allowing RFEA measurements every 50 μs. The capacitive effect from fast sweeping of the discriminating grid, which has previously limited this type of operation, is negligible in these experiments since the amplitude of the capacitance is lower than the signal noise. The other components in the RFEA (see [7] for more details) are held constant as with standard ion mode operation. The slit plate was biased at -160 V to repel electrons and allow a measurement of the ion saturation current ($I_{sat}$) on the slit plate and grid 2 was biased at -240 V for the type I ELM measurements and for type III ELM measurements at -220 V. This grid is used to supress secondary electrons resulting from ion impact on the back of the slit plate or the grids.

## 2.2 Slow Swept Technique (SST)

The SST is suitable for smaller ELMs such as type III ELMs since the ELM duration is smaller and therefore the assumption that the voltage is constant during the ELM holds. Also considerably less variation is seen between type III ELMs on MAST at the target [9]. The SST has been used in this work to provide a comparison between the SST and the FST measurements of type III ELMs in order to validate ELM measurements with RFEA probes. In this experiment the discriminating grid (grid 1) was swept at a rate of 40 Hz from 20 V to 200 V. The slit plate was held at -160 V and grid 2 was held at -220 V as with normal ion mode for RFEA measurements.

SST measurements have been made with the same analysis technique as used previously in [9] and for L-mode filaments measurements in [17]. Each ELM measurement provides a measure of $I_{sat}$ and the collector current ($I_{col}$) for specific discriminating voltage which can be combined to give the average ion temperature. The total ELM signal can be divided into periods of 50 μs for the analysis allowing the ion temperature through the average ELM to be determined in 50 μs intervals. The ion saturation current signal on the slit plate and the collector current are integrated in 50 μs intervals and then their ratio is taken to measure against the discriminating voltage.

## 2.3 RFEA Analysis

Both the SST and the FST technique determine $T_i$ by fitting equation (1) to a characteristic of the ratio of currents measured at the slit plate and collector ($I_{col}/I_{sat}$) against the discriminating voltage applied to grid 1 ($V_g$), see figure 2 for example characteristics from (a) the SST analysis and (b) the FST analysis. The ratio of the currents is used as opposed to solely the collector current in order to remove any density effects in the varying $I_{sat}$ signals incident on the divertor RFEA during an ELM. This is because a varying density at the probe would cause the measured current at the slit plate and collector to vary independently of $V_g$, therefore potentially giving an incorrect $T_i$ measurement. The characteristics used in this analysis use R = $I_{col}/I_{sat}$ against $V_g$. $I_{sat}$, as opposed to $j_{sat}$, is used in this analysis since the estimation of the wetted area of the slit plate, $A_{sp}$, during an ELM is subject to uncertainties and therefore this method allows the uncertainty in both $A_{sp}$ and the transmission through the RFEA probe, $\xi_r$, to be a fitted parameter.

$$\frac{I_{col}}{I_{sat}} = R = R_0 \exp\left(\frac{-(V_g - V_s)}{T_i}\right) + C \quad (1)$$

Where $R_0 = A_s \xi_r \xi_{opt}^3 / A_{sp}$ is the transmission factor for the probe (labelled as R=$I_{col}/I_{sat}$ for ratio as in [17]), where $A_s$ is the slit inlet area, $V_s$ is the sheath voltage, C accounts for any offsets in the current signals due to the electronics, and $\xi_r \xi_{opt}^3$ is the transmission through the probe which can be estimated using the probe dimensions. $\xi_r(v_{||})$ is a function of the parallel ion velocity because the proportion of ions which travel through the probe to the collector is dependent on their parallel energy due to the field alignment of the probe. An estimate of the total transmission has been made for the MAST divertor probe using the theory in [19], and is estimated to be in the range $\xi_r \xi_{opt}^3$ = 0.09 – 0.14.

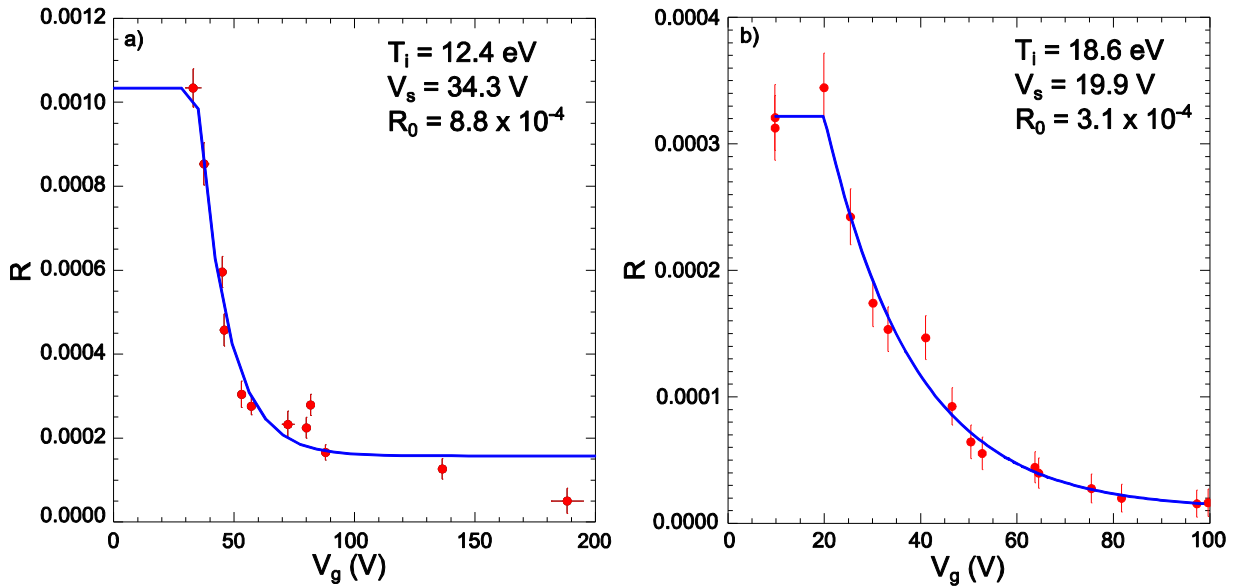

Figure 2 Example of a typical characteristic of R= $I_{col}/I_{sat}$ against $V_g$ for the a) SST and the b) FST analysis.

# 3 Experimental Results

## 3.1 Validation of ELM measurements by RFEA

ELM events are difficult to measure by electric probes at the edge of tokamaks because of the high heat loads and the fast timescales involved. Since RFEA probes rely on the assumption that the ion distribution measured is Maxwellian, measurements of ELMs can prove difficult. At the target, measurements of ELMs have been attempted by both the SST, as used on AUG [4] and MAST for L-mode filaments [17] and previous ELM measurements on MAST [17], and the FST. Here, two similar sets of type III ELMs are used to compare the two methods.

### *3.1.1 SST measurements of type III ELMs*

Previous measurements made using the SST have shown for an average type III ELM that $T_i$ decreases from an initial peak, soon after the ELM occurs, to the previously measured inter-ELM value [20] as a function of time [9]. The measurements here are in an equivalent DN discharge to that in [9] which produces type III ELMs. The measurements are taken at a position of $\Delta R_{LCFS}^{tgt}$ = 5.5 cm ± 0.5 cm.

Twelve similar ELMs, used in this analysis, were found by comparing their normalised $I_{sat}$, see figure 3.

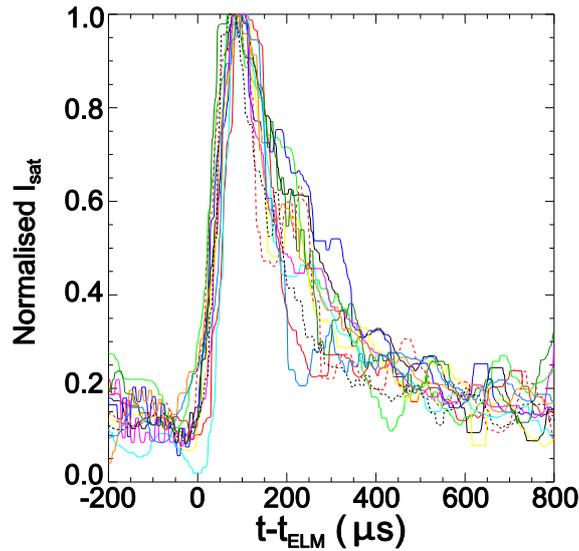

Figure 3 Twelve similar type III ELMs analysed by the slow sweep technique at a radial position of $\Delta R_{LCFS}^{tgt}$ = 5.5 ± 0.5 cm.

Using the same analysis technique as in [9], measurements of $T_i$, $V_s$ and $R_0$ are made as a function of time. Measurements of $T_i$, shown in figure 4, show a reduction in $T_i$ with t-$t_{ELM}$, similar to previous measurements [9], with $T_i$ falling to 5-10 eV. The peak $T_i$ is lower than previous measurements at $\Delta R_{LCFS}^{tgt}$ = 5 cm, but larger than those made at $\Delta R_{LCFS}^{tgt}$ = 7 cm [9], with $T_{i,peak}$ ~ 40 eV. $V_s$ shows a reduction with time, similar to previous measurements, ranging between $V_s$ = 50 V and 30 V. The electron temperature is related to $V_s$ by $V_s$ = 2.7$T_e$ [21] therefore this corresponds to an estimated $T_e$ = 20 – 10 eV which is consistent with steady state $T_e$ measurements although more elevated at the ELM start. This confirms the idea that the electrons from ELMs do not travel to as large a radius without cooling compared with the ions from ELMs. The large error bar on the final $V_s$ point in the range t-$t_{ELM}$ = 150 – 200 µs is due to the lack of measurements made in the range $V_g$ ≤ 20 V. This means that in this R-V characteristic only an upper limit for $V_s$ can be estimated and consequently a lower limit for $R_0$ which is reflected in the respective error bars on Figure 4(c). $R_0$ measurements can give some insight into the uncertain values of $A_{sp}$, since the wetted area isn't well defined during an ELM, and the transmission through the probe, $\xi_r(v_{||})$. The change in $R_0$ is likely related to a change of the wetted area through the ELM duration or a change in the transmission through the probe which is linked to the energy of the ions. The values measured here are consistent with calculations where a maximum of $A_{sp} = A_{geom}/3$, where $A_{geom}$ is the geometrical area of the slit plate, or a maximum transmission of $\xi_r = 0.95$. In reality the wetted area is likely to be lower than this but the change in $R_0$ as a function of time is most likely linked to the reduction in the ion energy which is also evident in the reducing $T_i$ values.

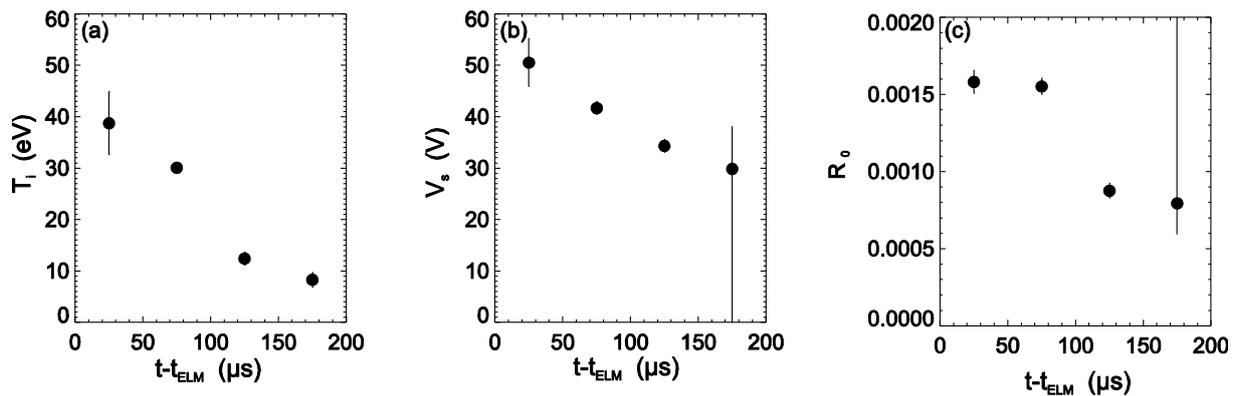

Figure 4 Time dependant results of (a) $T_i$, (b) $V_s$ and (c) $R_0$ from 12 type III ELMs measured at $\Delta R_{LCFS}^{tgt}$ = 5.5 ± 0.5 cm by the slow swept technique. Error bars are determined from the least squares fitting routine.

### 3.1.1 FST measurements of type III ELMs

The comparison set of type III ELMs for the FST analysis have been measured at the same radius relative to the plasma as the SST at $\Delta R_{LCFS}^{tgt}$ = 5.5 ± 0.5 cm. Since measurements are made continuously at a rate of 20 kHz with the FST, the resolution is much higher than the SST. To guide the eye for comparison the measurements in each 50 μs window have been boxcarred. Another advantage of the FST is that it allows measurements later in the type III ELM since when the signals are low in the SST technique it is difficult to fit the data with limited statistics. This is not the case with the FST since the sweeps are continuous so therefore the fitting is more robust. It should be noted that in the following data the $t-t_{ELM}$ axis extends to 300 μs rather than 200 μs as seen in the SST analysis.

A set of 26 similar type III ELMs have been measured by the FST and the normalised $I_{sat}$ signals can be seen in Figure 5.

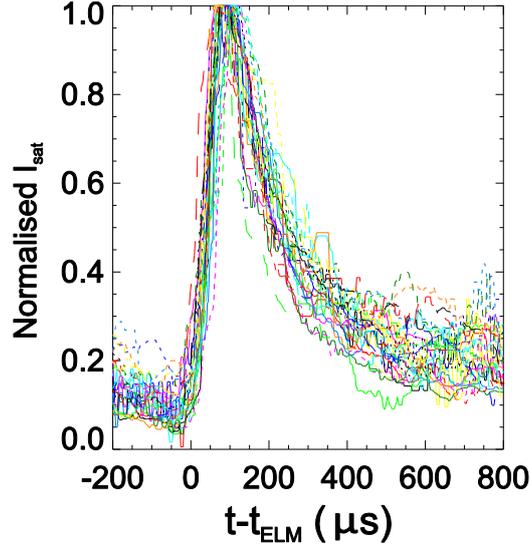

Figure 5 Set of 26 similar ELMs measured by the FST at $\Delta R_{LCFS}^{tgt}$ = 5.5 ± 0.5 cm which fit in a constrained window based on the normalised $I_{sat}$ measurements.

The fitted parameters of $T_i$, $V_s$ and $R_0$ for each sweep during a type III ELM have been plotted in figure 6. Data are colour coded by the time range in which the average time of the sweep falls and these ranges correspond to the time windows in the SST analysis. The boxcarred data in each time window is shown in black. Only sweeps after $t-t_{ELM}$ = 50 μs are shown since before this time the current at the collector is not sufficiently high to provide a reasonable fit to the R-V characteristic.

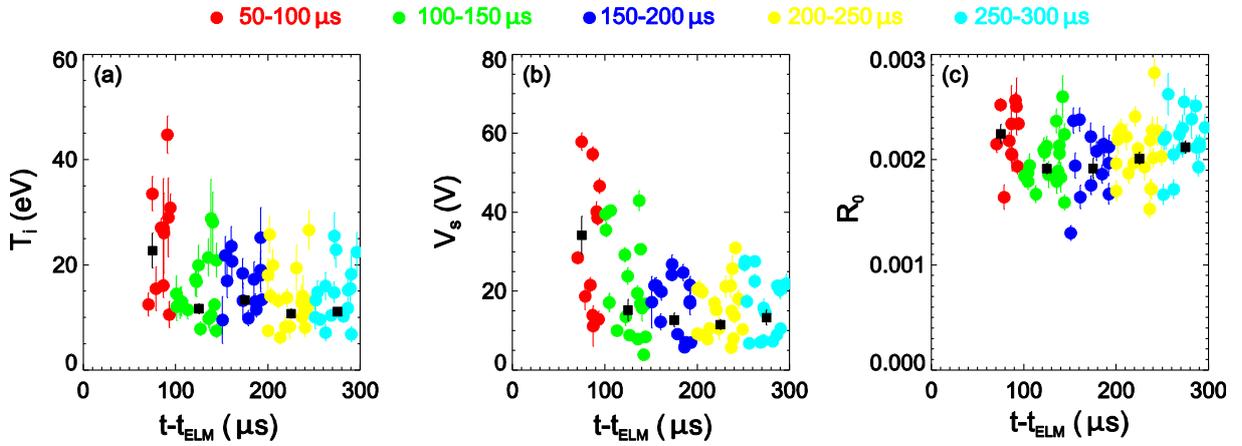

Figure 6 Time dependant results of $T_i$, $V_s$ and $R_0$ from 26 type III ELMs measured at $\Delta R_{LCFS}^{tgt}$ = 5.5 ± 0.5 cm using FST. Colours correspond to the time range in which the sweeps fall. Error bars are determined from the least squares fitting routine for the FST points.

The FST data for $T_i$ shows some agreement with the SST however it shows that the SST measurements are dominated by the hottest ions because the SST overlay with the highest FST measurements. The FST data scatter suggests that the density effects may not be entirely removed by taking the ratio of the currents. This is anticipated since there is a slightly weaker dependence on temperature in the current, as well as density. A general reduction of $T_i$ during the ELM evolution is seen as expected from previous measurements [9]. There is possibly a finer structure in temperature in the type III ELMs than is easily investigated by this method since type III ELM signals exist on the RFEA for short timescales compared to type I ELMs which can be investigated in more detail. The general trend of $T_i$ and the agreement within a factor of ~2-3 confirms that the methods are sufficiently consistent to use the FST for detailed investigation. Ideally to investigate type III ELMs faster sweep times would be used, however these are not possible with the current electronic sweep rate capabilities. As with $T_i$, $V_s$ has a similar trend to the SST data however the scatter in the FST shows that the SST only had the capability to show the upper limit since the fits appear dominated by the highest energy ions and electrons. The values of $V_s$ measured which are at the low end of the scatter show very low $T_e$ estimates which isn't unlikely since inter-ELM little or no current signal is seen on the RFEA and therefore the electron temperatures in this region are expected to be cold. Measurements of $R_0$ are relatively constant and slightly higher than measured by the SST method. This is likely due to an underestimate in the SST fitting since the $T_i$ and $V_s$ are biased by the hottest ELM ion measurements which would therefore underestimate $R_0$. Particularly since measurements below $V_g$ = 20 V are unavailable it would be wise to favour the FST measurements.

## 3.2 Type I ELM observations at the divertor

Measurements of type I ELMs at the MAST divertor have shown varying structures in the current arriving at the RFEA slit plate compared to type III ELMs, as seen on AUG [22], which is likely linked to the greater distance these measurements are made from the strike point at the target ($\Delta R_{LCFS}^{tgt}$). Such observations make it difficult to determine 'similar' type I ELMs for the SST measurements of an average type I ELM and therefore they are an ideal candidate for the new measurement technique.

It is likely that, since the measurements are made at relatively large $\Delta R_{LCFS}^{tgt}$ positions for type I ELMs, the RFEA measures filamentary parts of type I ELMs which are radially distant from the peak ELM heat flux. They are of particular interest since these filaments are arriving in regions where high heat flux and particle flux is not expected and therefore it is important to assess the temperatures of the ions arriving.

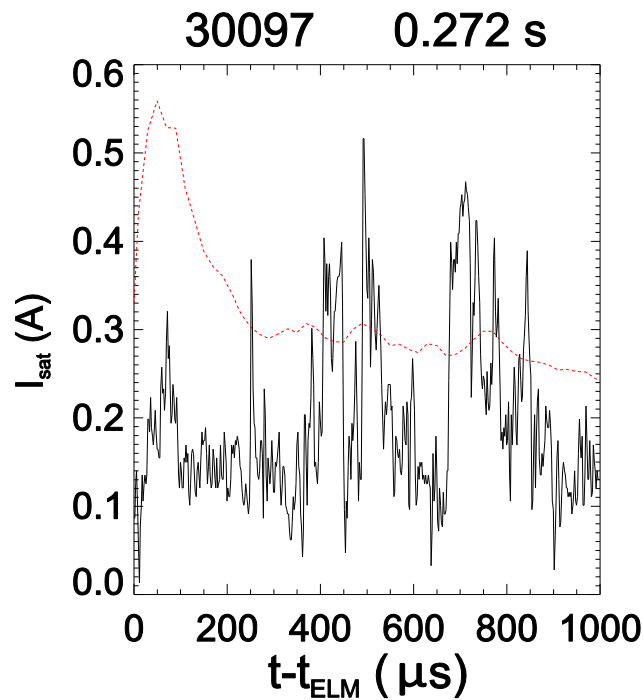

**Figure 7 $I_{sat}$ at divertor RFEA during a type I ELM event measured when the strike point is more than 15 cm away from the RFEA probe with $D_\alpha$ at the midplane shown in red.**

Filaments have been routinely seen as stripes on the divertor on many tokamaks by various camera measurements [13] [23] [24]. Figure 7 shows the variation in $I_{sat}$ seen at the RFEA slit plate in a type I ELM at $\Delta R_{LCFS}^{tgt} > 15$ cm as a function of time along with $D_\alpha$ at the mid-plane. IR thermography measurements have been used to confirm that measurements seen on the RFEA $I_{sat}$ signal are related to type I ELM filaments. The IR camera views the outer divertor of MAST one toroidal sector from the RFEA location which corresponds to an offset of 22.5° toroidally and a time difference of 2 – 3 µs when measuring an ELM filament. IR measurements have been found to show individual filaments which correlate with ELMs [13]. In this particular plasma the time resolution on the IR camera isn't sufficient to allow filament identification at the radius of the RFEA probe but the link with the type I ELM can be confirmed. Figure 8(b) shows the heat flux, measured by IR, at the RFEA location (R = 0.9855 m) as black dots. An increase in heat flux is seen at the same time as the $I_{sat}$ signal is seen on the RFEA, figure 8(c), and both are synchronised with the $D_\alpha$ signal showing the type I ELM. Looking at the heat flux as a contour plot around the RFEA location in the range R = 0.9 – 0.99 m, there is an isolated peak in heat flux which suggests an individual filament has arrived close to the RFEA position although this is later in time than the signal seen on the RFEA. Therefore it seems that the signal seen on the RFEA is synchronised with the type I ELM although it appears to be made out of several small filaments compared to the peak heat flux seen on the IR. It is as yet unclear, due to the resolution of the IR camera in this plasma, what type of filament is creating the structures seen on the RFEA $I_{sat}$. Although it seems as though the IR sees the increase in heat flux after the RFEA measures the ELM associated current and the increase in mid-plane $D_\alpha$ (see figure 8(a)), this is due to the integration time of the IR camera (280 µs) which is shown in figure 8(c) by the grey panel. This suggests that the structures seen in the RFEA $I_{sat}$ are present in the IR heat flux measurement but are not clear due to the lower time resolution in the IR measurements for this particular plasma.

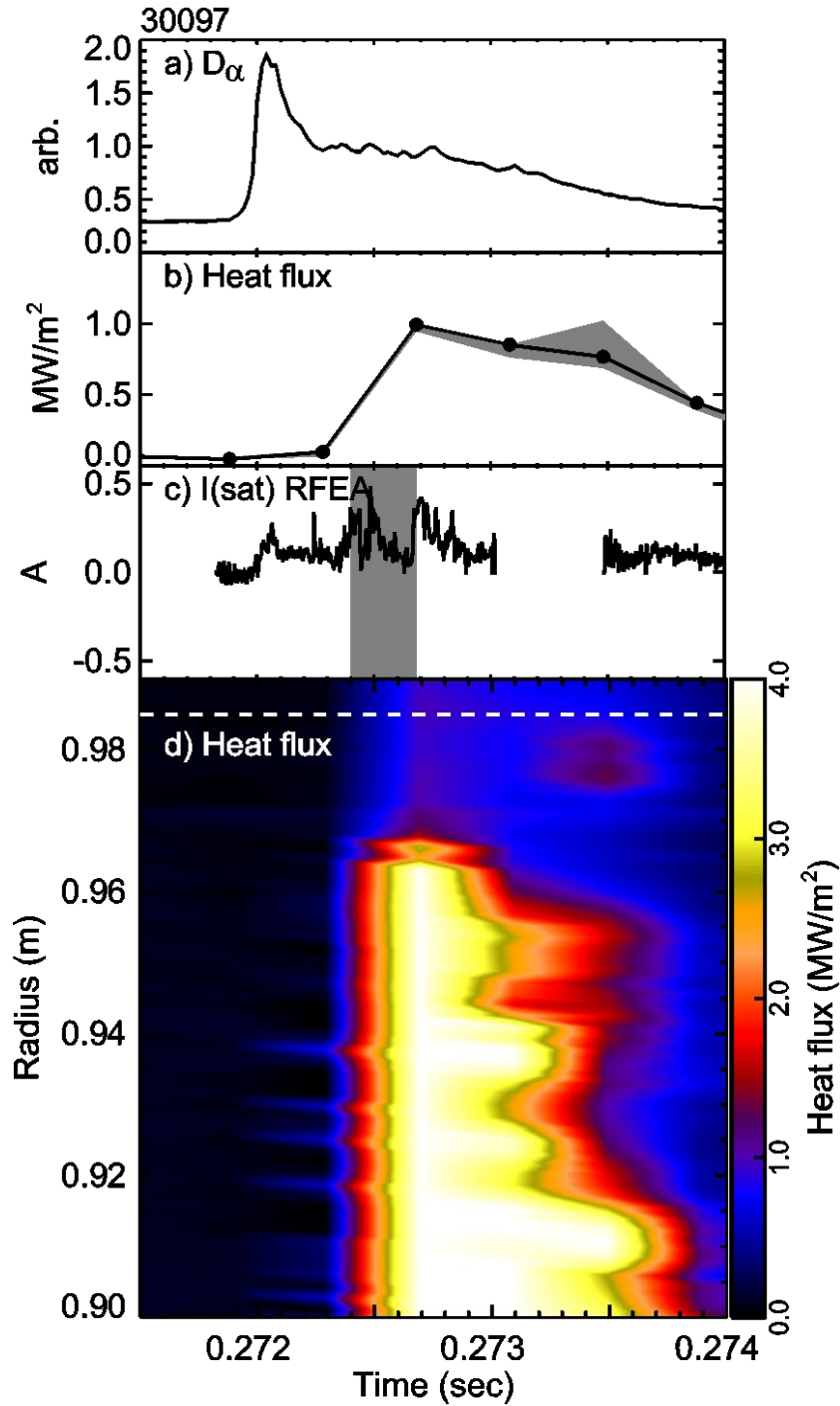

Figure 8 Comparison of IR thermography measurements in a region of the lower divertor close to the RFEA location showing (a) $D_\alpha$, (b) IR heat flux, (c) RFEA $I_{sat}$ and (d) contour heat flux plot from the IR during the same type I ELM event seen in figure 7. The horizontal white line in (d) shows the RFEA radial location. Gaps in data are due to PF noise on the RFEA signals.

In order to establish whether the variation in current with increasing $t-t_{ELM}$ is related to parallel transport along filaments or filament movement past the probe, the $I_{sat}$ signal can be compared to transport timescales. It has been found that ELMs have a toroidal velocity of $v_\phi$ = 10 - 30 kms$^{-1}$ [12], which decreases when the ELM filament separates from the plasma [13]. A Monte Carlo code to simulate the target particle flux from ion parallel transport [13] has been used to model the time taken for a type I ELM filament to reach the radial location of the RFEA at the target, $\Delta R_{LCFS}^{tgt} \geq 15$ cm.

The particles in the simulation reach the detector at t-$t_{ELM}$ = 120 µs. Using deceleration estimates for the toroidal rotation of filaments from [13] it can be found that by the time the filament reaches the RFEA it will not be moving significantly in the toroidal direction during the lifetime of the ELM $I_{sat}$ signal. The radial motion however is significant by the time the particles reach the RFEA since the radial velocity, $v_r$ = 2 – 9 kms$^{-1}$ [12], increases after the filament separates from the plasma [13]. To find if the radial motion would cause the structures found in the measured $I_{sat}$ at the RFEA, the maximum time a filament would be seen at the RFEA is estimated. The radial acceleration for ELM filaments has been estimated as 0.7 – 2.5 x 10$^8$ ms$^{-2}$ [13]. Taking the minimum speed and acceleration, the slowest the filament can be moving at $\Delta R_{LCFS}^{tgt}$= 15 cm is $v_r$ = 10.4 kms$^{-1}$. Using the probe aperture width, w = 2 mm, which allows access to the slit plate and the estimated size of a filament at the mid-plane ($L_r$ = 2 – 6 cm [12]) expanded in flux for the plasmas we are looking at, with flux expansion $f_E$ ~ 5; the maximum distance the filament can extend at the RFEA location is 30.2 cm. This gives an estimate of ~ 35 µs for the signal from a filament to exist on the RFEA slit plate. This is of order of the peaks seen in $I_{sat}$, however since there is an $I_{sat}$ signal seen at the probe for t-$t_{ELM}$ > 1 ms; it is likely that the signal is made up of several filaments passing the RFEA probe, particularly as the slit width used here is likely an overestimate of the wetted area of the RFEA slit plate in the radial direction.

## 3.3 FST Type I ELM $T_i$ Measurements

Type I ELMs have been measured in the range t-$t_{ELM}$=0 – 1 ms; this is a larger range than in type III ELMs because the type I ELM current signal exists longer at the divertor RFEA. The most likely reason for this is that type I ELMs have larger footprints than type III ELMs at the divertor [25] and therefore if the ELM moves past the divertor RFEA then signal will be seen for a longer time. As has already been shown, each type I ELM measured has a different current structure measurement which appears to be several filaments moving past the probe radially. The following measurements show $T_i$, $V_s$ and $R_0$ as a function of t-$t_{ELM}$ which have been determined by fitting the R-V characteristics. Any gaps seen in the data is due to pick up noise from the MAST poloidal field (PF) coil switching; during these periods data were not used for analysis since the PF coil switching induces currents in the measurements.

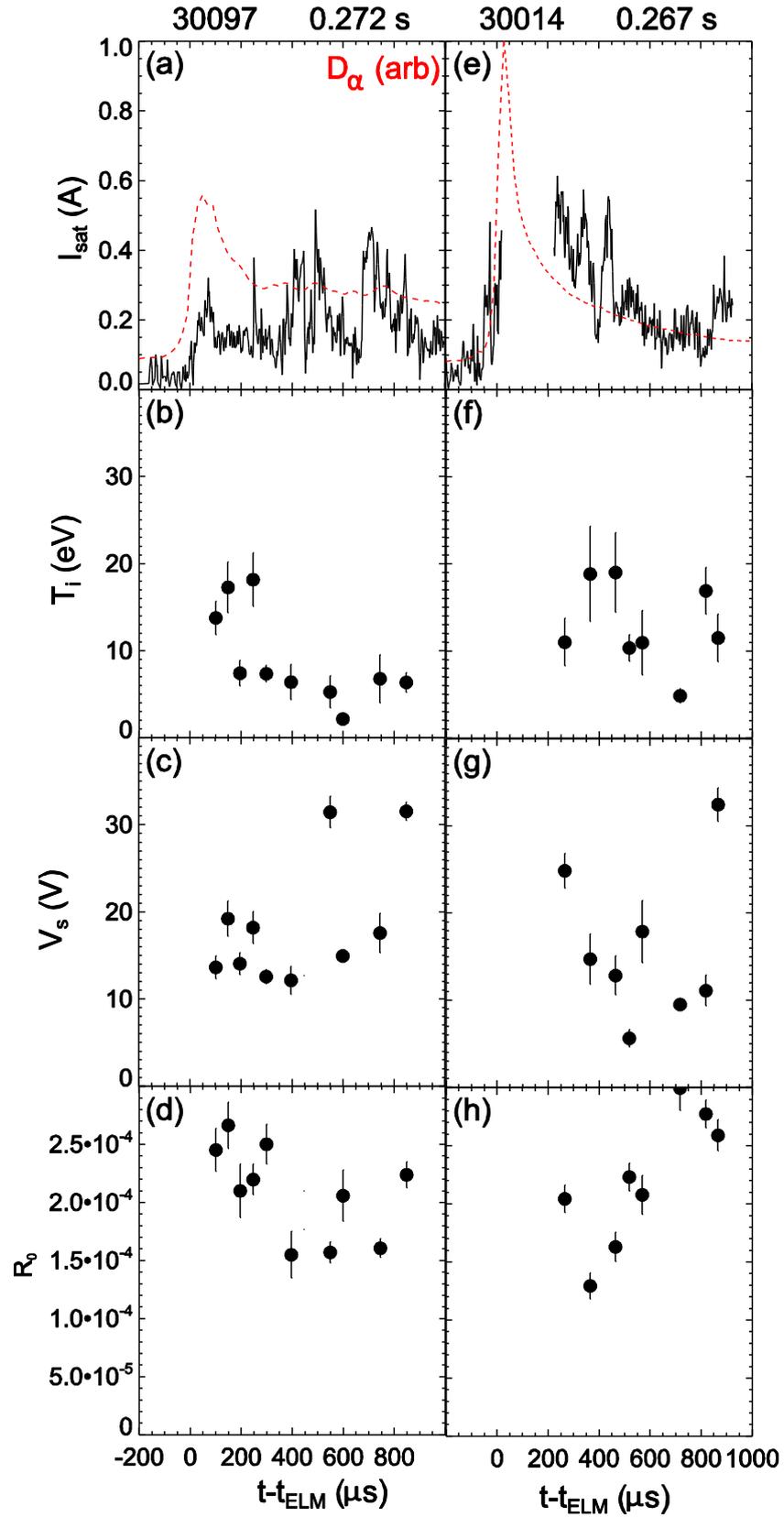

Figure 9 $I_{sat}$, $T_i$, $V_s$ and $R_0$ as a function of $t-t_{ELM}$ for two type I ELM at the outer lower divertor as measured by FST. Error bars are determined from the least squares fitting routine.

Figure 9 shows two type I ELM measurements of $I_{sat}$, mid-plane $D_\alpha$, $T_i$, $V_s$ and $R_0$; the first ELM is the one studied in figure 7 and figure 8 and the second is from a repeated plasma. Both ELMs show temperatures of up to 20 eV which proves temperatures comparable to the steady state strike point are found more than 15 cm from the strike point due to ELMs. The same trend of decreasing temperature with t-$t_{ELM}$ is found as with the type III ELMs which shows that the ions lose energy after the initial filament separation and is evidence that the later filaments are cooler generally than the ones arriving at the target first following the ELM event. Although in the second ELM the period around t = $t_{ELM}$ is unavailable due to PF coil pick up noise on the RFEA, temperature peaks appear in the range t-$t_{ELM}$ = 300 – 500 μs which appear to be synchronised with the peaks seen in $I_{sat}$. This is in contrast to the peaks in $I_{sat}$ seen after 400 μs in the first ELM where no ion temperature increase is seen to synchronise with the $I_{sat}$ increase. This suggests that we may have two different types of filaments arriving at the DSF RFEA; hot primary filaments released from the ELM and colder secondary filaments.

$V_s$ does not follow the expected trend from the type III measurements of a slow decay with t-$t_{ELM}$ to a value equivalent to steady state $T_e$ for either of the type I ELMs shown here. Since $V_s$ = 2.7$T_e$ the measurements at this radius are expected to be low since there should be few inter-ELM electrons present. The lower values in the first ELM are equivalent to $T_e$ ~ 5 eV or lower which is possible at this large radius although a significant density of electrons is not expected. The peaks in $V_s$ at 550 μs and 850 μs are unlikely to be associated with inter-ELM electrons and are therefore most likely linked to the increases in $I_{sat}$ at these times. It is unclear why the electrons seem to have higher energy here than the ions, however these measurements are only equivalent to $T_e$ ~ 12 eV and therefore the electrons are considerably cooled compared to the expected temperature of ELM electrons arriving closer to the strike point. The second ELM studied does show the expected decay of $V_s$ with t-$t_{ELM}$ and again the maximum temperatures seen here for the electrons are of order $T_e$ ~ 12 eV which is not significant although it is unusual at this large radius. The peak in $V_s$ later in time at t-$t_{ELM}$ = 850 μs is consistent with the most significant $T_e$ measured in these ELMs which is surprising so late after the ELM occurs. This increase could be associated with the start of an increase in $I_{sat}$ at the same time. This is likely a primary filament moving past the probe later in time in relation to the initial ELM event and is synchronised with the later increase seen in $T_i$.

$R_0$ is between 1 x $10^{-4}$ and 3 x $10^{-4}$ which is a factor of 10 lower than type III ELM measurements by FST. This is either due to a large reduction in the parallel transmission or an increase in the wetted area of the RFEA slit plate during type I ELMs at larger radius. These contributions compound the effect as at large radius the weak B field will mean that ions have large Larmor radii and therefore the transmission will be reduced through the probe but the wetted area will potentially be increased above the actual geometrical size of the probe aperture. This hypothesis is consistent with calculations of the ratio of $A_{sp}/\xi_r$ which give an estimate of 106% of $A_{sp}$ when the average estimated value for the MAST RFEA of $\xi_r$ ~ 0.2 is used showing that the wetted area would need to extend beyond the area of the inlet slit. Alternatively, lower transmissions of $\xi_r$ ~ 0.19 and below would be consistent with an area of $A_{geom}$ when using the lowest measurements of $R_0$. In the first ELM shown in Figure 9 shows a decrease of $R_0$ which is likely linked to the ion energy since $T_i$ is seen to decrease over a similar time. The second ELM shows a variation that isn't likely to be linked to the ion energy. It is more likely linked to a reduction in the wetted area of the probe as the ELM filaments move across the RFEA probe inlet, however this cannot be thoroughly investigated for this ELM since there is no visual data available on the probe with sufficient resolution.

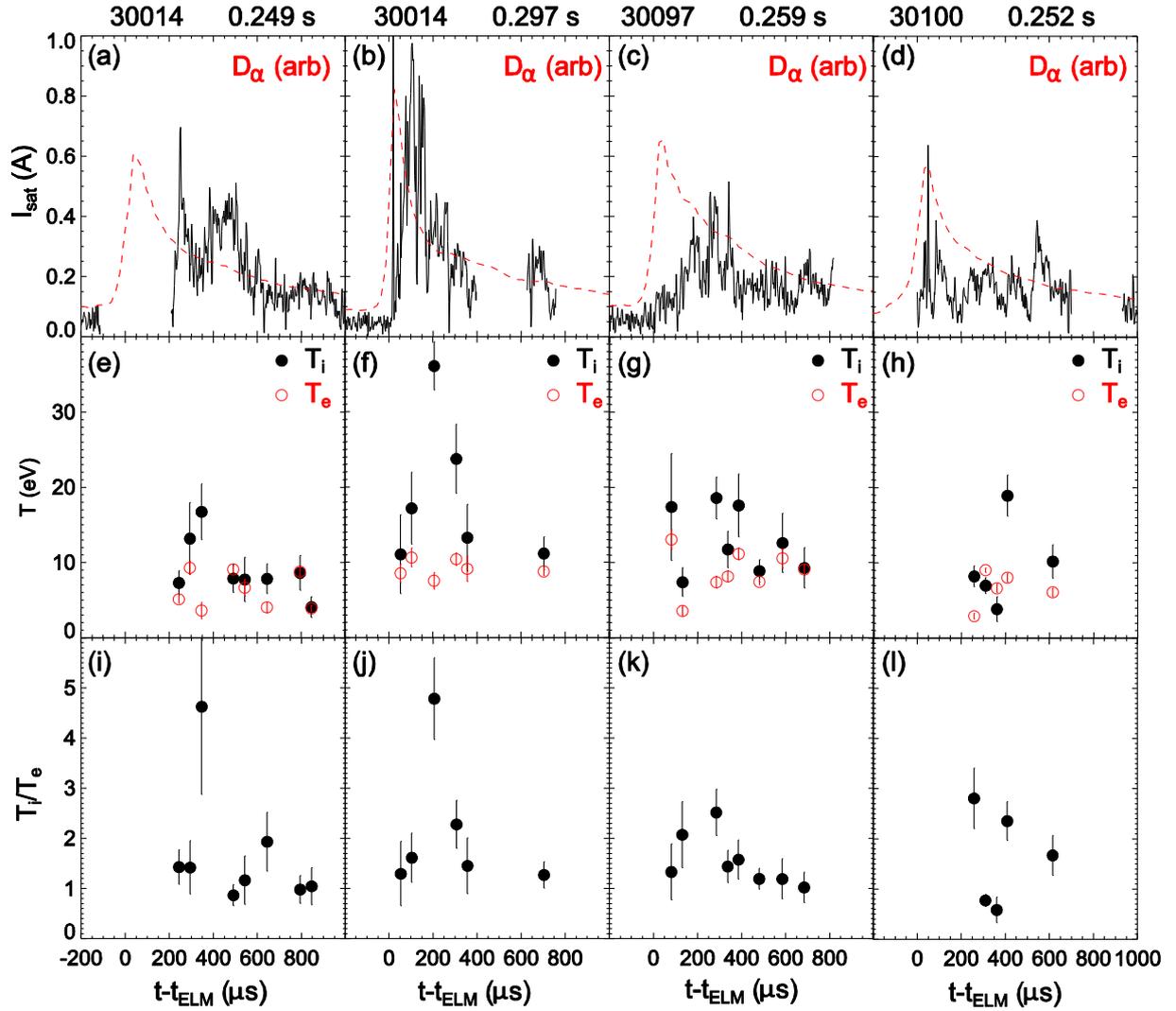

Figure 10 $I_{sat}$, $D_\alpha$, $T_i$, $T_e$, and $T_i/T_e$ as a function of $t-t_{ELM}$ for four type I ELMs at the outer lower divertor as measured by RFEA. Error bars are determined from the least squares fitting routine.

Figure 10 shows measurements of $I_{sat}$, mid-plane $D_\alpha$, $T_i$, $T_e$ and the ratio of $T_i/T_e$ for four further type I ELMs. The $I_{sat}$ and mid-plane $D_\alpha$ traces can be seen in Figure 10(a-d) and show that for each mid-plane $D_\alpha$ trace with similar evolution, the $I_{sat}$ signal varies considerably with $t-t_{ELM}$ at the target. In Figure 10(e-h) $T_i$ and $T_e$, estimated by $T_e = V_s/2.7$ [21], are shown. Generally $T_i$ varies in the range $T_i = 5 - 20$ eV with some outliers and $T_e$ is systematically lower with $T_e < 15$ eV. The ratio of $T_i/T_e$ is shown in Figure 10(i-l) and generally lies in the range $T_i/T_e = 1 - 2.5$.

The evolution of $T_i$ is different for each ELM and since the structures seen in $I_{sat}$ are thought to be plasma moving past the probe, the evolution of temperature can give some idea of where the filaments are coming from. Generally $T_e$ is low suggesting that hot electrons from the ELM do not reach large radius. The values for $T_e$ are still higher than are expected at this radius, where $T_e$ is not usually measured, which suggests that either electrons remain up to $T_e \sim 10$ eV far from the LCFS but that, in MAST, the current signals are too low to easily measure by LP, or that ions released by the ELM have a heating effect on the local cold electron population. This is possible since given MAST conditions where we assume $n_e \sim 1\times10^{17}$ m$^{-3}$, $T_e < 5$ eV and $T_i > 25$ eV; $T_e$ can increase to 10 eV within $\sim 300$ μs given the ion-electron collision time [17]. Where peaks in $T_i$ correspond to a peak in $I_{sat}$ these ions are most likely resulting from primary ELM filaments, for example see Figure 10(a,b,e,f). This effect is also seen in Figure 10(c,g) between $t-t_{ELM}$ = 200 – 400 µs. This particular ELM is interesting because there is also a peak in $T_i$ at the beginning of the ELM where there is not a significant rise in $I_{sat}$ seen. This would suggest high temperature ions arriving with a low density, which is likely the reason that the error bar is slightly larger. In the cases where filament signals are seen in the $I_{sat}$ trace but there is not significant $T_i$ increase, generally later in time relative to the ELM, it is thought that these are due to secondary ELM filaments with lower $T_i$, occurring as a result of the primary ELM filaments. This can be seen in the ELM shown in Figure 10(b,f) after $t-t_{ELM}$ = 600 µs and also in the majority of the measurements made in the ELM shown in Figure 10(d,h). A discussion of the presence of secondary filaments will be explored in the following section.

The ratio of $T_i/T_e$ during ELMs has implications for interpretive measurements made at the divertor which assume $T_i = T_e$, such as electron density, $n_e$, from LPs and power to the divertor, $P_{div}$, which relies on an assumption for the heat transmission factor, γ. Here it is seen that $T_i \geq T_e$ and sometimes the ion temperature is significantly higher; with $T_i/T_e$ = 2.5 regularly seen and excursions up to a factor 4. This means that density at the target could be overestimated and heat flux, related to $P_{div}$, underestimated; both of which would lead to insufficient divertor protection in these regions where high heat loads and temperatures are not normally expected. The other interesting fact is that this confirms that although electrons released from ELMs deposit their energy closer to the strike point, ions can travel much further with their energy to regions which aren't expecting any heat loads as seen in [5] and [2].

## 3.4 Secondary ELM filaments

Type I ELM signals, seen on the RFEA slit plate, continue for extended times, >1 ms, which is not normally seen on MAST [18]. It is postulated that the temperature and $I_{sat}$ fluctuations seen in the RFEA measurements are the result of filaments moving past the RFEA probe radially due to estimates of the timescales on which they occur. It is unlikely that these are all primary type I ELM filaments since, at locations of $\Delta R_{LCFS}^{tgt} \geq 15$ cm, the spiral structure of filaments at the target [25] means the probe is unlikely to be passed by several primary filaments in every ELM during the 1 ms interval that has been studied. Also, although the levels of $I_{sat}$ and $T_i$ are significant compared to inter-ELM signals, where almost no signal can be seen, measured signals during the ELM are lower in amplitude compared to measurements of large type I ELMs made at closer $\Delta R_{LCFS}^{tgt}$ positions which can often saturate the RFEA power supplies. These observations support the notion that some of the measurements studied here, at $\Delta R_{LCFS}^{tgt} > 15$ cm, are measurements of what are known as secondary filaments.

Secondary filaments have been seen on a number of tokamaks such as NSTX, JET and AUG [10] [26] [27] [28]. Previously secondary filaments have not been identified in MAST [28] however this is postulated to be related to the remote wall in MAST since secondary filaments have been found to occur when primary ELM filaments interact with plasma facing structures.

On MAST, the duration of signal seen following an ELM varies with the magnetic configuration. Figure 11 shows data from both mid-plane $D_\alpha$ and $I_{sat}$ traces on the Ball Pen Probe (BPP) [14] on the mid-plane reciprocating probe for two type I ELMs in different magnetic configurations. The two magnetic configurations, LSN and DN, are also operated at different Z positions; as can be seen in Figure 1, the LSN in MAST is operated closer to the P3 coil structure. It can be seen in Figure 11 that both the $D_\alpha$ and also $I_{sat}$ signals measured on the BPP pins for LSN exist for much longer than an equivalent type I ELM in DN configuration. These extended signals are unlikely to be primary filaments and are expected to be secondary filaments which are not seen in standard DN ELMy H-mode plasmas.

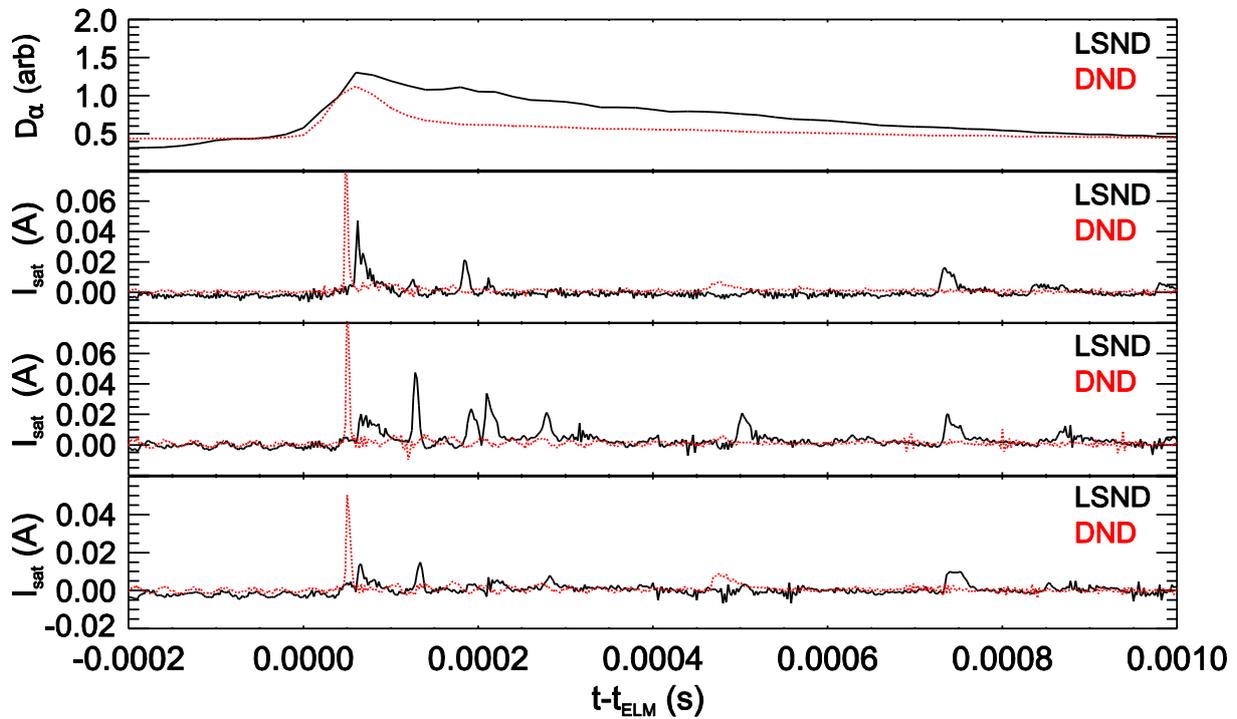

Figure 11 $D_\alpha$ and BPP data at the mid-plane for an LSN and a DN plasma during type I ELMs

Evidence of the extended filamentary behaviour in LSN compared to DN plasmas can be also seen on fast camera images, see Figure 12. Panels (a)-(f) show the evolution of a type I ELM in LSN every 200 µs from $t_{ELM}$-200 µs to $t_{ELM}$+800 µs. Filamentary behaviour can be seen at $t_{ELM}$, primary ELM filaments, and then the activity is lower but still present beyond $t_{ELM}$, with a large interaction seen with the P3 coil, particularly at $t_{ELM}$+200 µs. The corresponding time windows following a type I ELM in DN configuration, panels (g)-(l), shows the same primary filaments at $t_{ELM}$ but no following filamentary activity is seen. Also it should be noted that there is no strong interaction seen with PFCs.

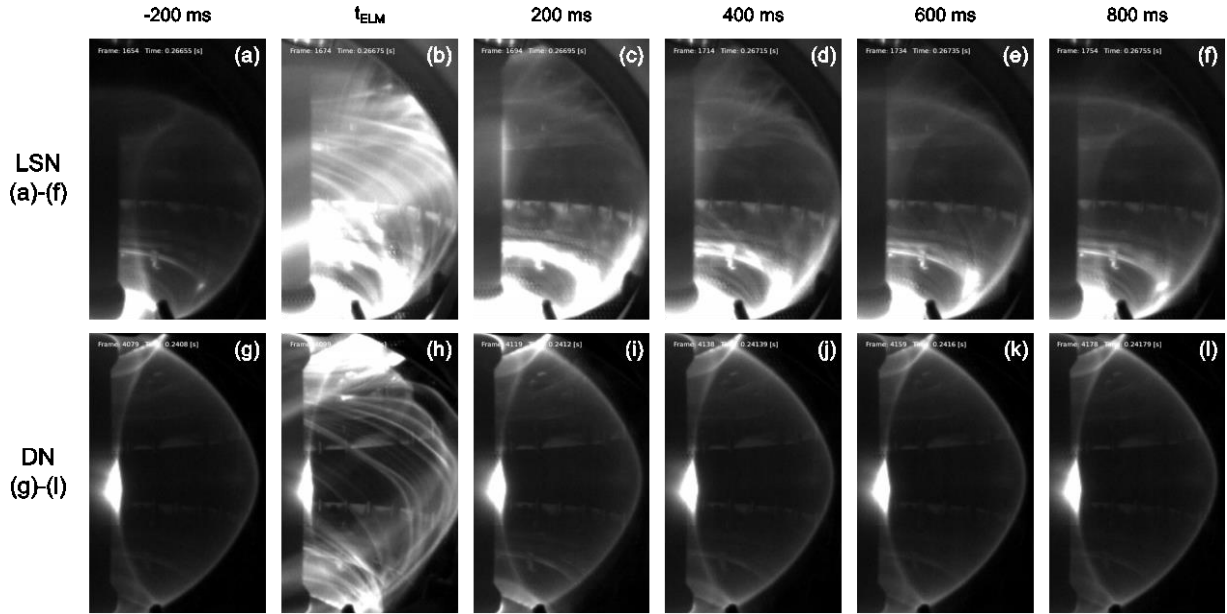

Figure 12 Fast Camera images for LSN and DN plasmas during type I ELMs

Type I ELMs studied in this work with the divertor RFEA occur during an LSN plasma. It is postulated that primary filaments from the type I ELM have caused secondary ELM filaments, due to interaction with the P3 coil structures, and this is what is measured by the divertor RFEA later in t-$t_{ELM}$. This would help to explain the oscillations in $T_i$ measured as a function of t-$t_{ELM}$ following the initial decay of temperature and ion flux following the assumed primary ELM filaments which are measured at t=$t_{ELM}$.

# 4 Conclusions

A new technique, the fast swept technique (FST), has been used to analyse both type III and type I ELMs following the comparison of the method with measurements using the SST [9]. Increases in $I_{sat}$, due to type I ELMs, are measured on the target RFEA at least 15 cm from the strike point. In this region there is no inter-ELM signal and therefore we can assume no significant amount of background ions reach this radius during inter-ELM steady state. Nevertheless, ion temperatures of up to 20 eV are routinely found during ELM events and some measurements of $T_i$ ~ 30 eV have also been seen. The highest measured inter-ELM $T_i$ in an equivalent plasma was $T_i$ = 8 ± 2 eV at $\Delta R_{LCFS}^{tgt}$ = 10 cm [20]. By comparing measurements with the IR diagnostic it has been found that it is most likely that a series of filaments have impacted on the target RFEA following an ELM event, any variations of $T_i$ during the $I_{sat}$ signal measured at the time of the ELM are likely to be related to filament structures passing the RFEA probe radially due to the timescale of the variations. Filaments at this location and time are possibly secondary ELM filaments resulting from primary ELM filament interaction with the lower P3 coil structures in MAST. Although secondary filaments have not been identified previously on MAST, the scenario studied here for type I ELMs is an LSN plasma which is operated closer to the lower P3 coil structures than any DN MAST plasma. Both measurements of $I_{sat}$ by BPP at the mid-plane and fast camera imaging have shown the presence of low amplitude filaments existing in LSN plasmas but not in DN which supports the hypothesis that some of the $T_i$ measurements made during type I ELMs are of secondary ELM filaments.

## Acknowledgements

This work was funded by the RCUK Energy Programme [under grant EP/I501045] and the European Communities under the contract of Association between EURATOM and CCFE. To obtain further information on the data and models underlying this paper please contact PublicationsManager@ccfe.ac.uk. The views and opinions expressed herein do not necessarily reflect those of the European Commission."